\documentclass{optica-article}
\journal{opticajournal} % for journals or Optica Open
\usepackage{xspace} 
\usepackage{gensymb}
\usepackage{graphicx}
\usepackage{caption}
\usepackage{subcaption}
\usepackage{url}
\usepackage{comment}
\usepackage{hyperref}
\usepackage{tabularx}
\usepackage{amsmath, amsfonts}
\usepackage{lineno}

\newcommand{\etal}{\textit{et al.\@}\xspace}
\newcommand{\exvivo}{\textit{ex vivo}\xspace}

\newcommand{\invivo}{\textit{in vivo}\xspace}
\newcommand{\Invivo}{\textit{In vivo}\xspace}
\newcommand{\invitro}{\textit{in vitro}\xspace}

\newcommand{\enface}{\textit{en face}\xspace}
\newcommand{\Enface}{\textit{En face}\xspace}
\newcommand{\um}{\(\muup\)m\xspace}

\newcommand{\deltaz}{{\Delta z}\xspace}

\newcommand{\ftcal}[1]{\mathcal{F}\left[{#1}\right]\xspace}

\begin{document}

\title{Multi-focus averaging for multiple scattering suppression in optical coherence tomography}

\author{
Lida Zhu,\authormark{1} 
Shuichi Makita,\authormark{1}
Junya Tamaoki,\authormark{3} 
Antonia Lichtenegger,\authormark{1,2} 
Yiheng Lim,\authormark{1} 
Yiqiang Zhu, \authormark{1}
Makoto Kobayashi,\authormark{3} 
and Yoshiaki Yasuno\authormark{1,*}}

\address{\authormark{1}Computational Optics Group, University of Tsukuba, Tsukuba, Ibaraki, Japan\\}
\address{\authormark{2}Center for Medical Physics and Biomedical Engineering, Medical University of Vienna, Vienna, Austria\\}
\address{\authormark{3}Department of Molecular and Developmental Biology, Institute of Medicine, University of Tsukuba, Japan\\}
\email{\authormark{*}yoshiaki.yasuno@cog-labs.org} %% email address is required
\homepage{https://optics.bk.tsukuba.ac.jp/COG/} %% author's URL, if desired

\begin{abstract}
Multiple scattering is one of the main factors that limits the penetration depth of optical coherence tomography (OCT) in scattering samples. 
We propose a method termed multi-focus averaging (MFA) to suppress the multiple-scattering signals and improve the image contrast of OCT in deep regions.
The MFA method captures multiple OCT volumes with various focal positions and averages them in complex form after correcting the varying defocus through computational refocusing.
Because the multiple-scattering takes different trajectories among the different focal position configurations, this averaging suppresses the multiple-scattering signal.
Meanwhile, the single-scattering takes a consistent trajectory regardless of the focal position configuration and is not suppressed. 
Hence, the MFA method improves the signal ratio between the single- and multiple-scattering signals and improves the image contrast.
A scattering phantom and a postmortem zebrafish were measured for validation of the proposed method. 
The results showed that the contrast of intensity images of both the phantom and zebrafish were improved using the MFA method, such that they were better than the contrast provided by the standard complex averaging method. 
The MFA method provides a cost-effective solution for contrast enhancement through multiple-scattering reduction in tissue imaging using OCT systems. 
\end{abstract}

%%%%%%%%%%%%%%%%%%%%%%%%%%  body  %%%%%%%%%%%%%%%%%%%%%%%%%%
\section{Introduction}
Deep tissue imaging has long been of strong interest in biomedical optics\cite{Gigan_2017}. 
Optical coherence tomography (OCT) provides three-dimensional images of a sample with high penetration and high resolution, and has become widely used for the non-invasive imaging of biological samples. %\cite{kagemann_MV_2008, marschall_ABC_2011}. 
OCT has been successfully adapted to investigations of various tissues, such as human skin\cite{Olsen_2018}, \exvivo brain tissue\cite{Kut_2015, Lichtenegger_2018}, and \invitro tumor spheroids\cite{Huang2017CancRes, Abd_El_Sadek_2020, ElSadek2021BOE}. 

Typically, an OCT image is mainly reconstructed from the single-back-scattering (SS) signals of the sample. 
The SS signals are retrieved from the photons that have been back-scattered once at the sample object and hence directly carry the object information. 
However, when imaging in scattering samples such as tissues, the multiple-scattering (MS) signals harm the image contrast especially in deep regions\cite{yadlowsky_AOA_1995}.
The MS-originated signals at a position in the image are not scattered at the corresponding position in the sample and thus do not convey the correct object information. 
MS limits the resolvable imaging depth of OCT in tissues and hampers the visualization of deep microstructures\cite{dunsby_JPDAP_2003, boas__2016}.

In standard OCT with a confocal configuration, most of the MS signals can be rejected by the confocal gating.  
However, the residual MS signals still disturb the imaging. %\cite{yadlowsky_AOA_1995}. 
Techniques for further reducing the MS effect have been explored. 
One approach is modulating the probing light to decorrelate the MS signals using wavefront manipulation devices, such as a spatial light modulator (SLM) \cite{badon_SA_2016, Wojtkowski_stoc2019} or a deformable mirror \cite{liu_BOEB_2018a}. 
However, this approach may result in high system complexity and cost. 
A solution with inexpensive optics that can be easily implemented in the standard OCT scheme is more preferable.

In this paper, we propose a method termed ``multi-focus averaging'' (MFA),to reduce the MS signals and improve the image contrast in deep regions of scattering samples. 
The MFA method uses a low-cost electrical tunable lens to decorrelate the MS signals over multiple volumetric acquisitions.
Here, the depth position of the focus is modulated among the multiple volumes. 
A computational refocusing is then adopted to cancel out the different defocus in the focus-modulated OCT volumes. 
A scattering phantom and a postmortem zebrafish were used for validation. 
The phantom results show that the MFA method achieved a 4.5-dB enhancement of the signal-to-background ratio (SBR). 
The postmortem zebrafish results show that the MFA method better visualized anatomic structures in deep regions, suggesting that this method can be used for biological imaging. 

\section{Principle and core methods}
\subsection{Principle of the MFA method}
\label{sec:principle:theory}

The MFA method is a combination of sequential volumetric OCT imaging with different defocus, computational refocusing of each volume, and complex averaging of the volumes.
In an OCT volume of a scattering sample, the complex \enface OCT signal $S(x,y;z, z_d)$ at an imaging depth of $z$ comprises two components, namely $S_{SS}(x,y;z,z_d)$ and $S_{MS}(x,y;z,z_d)$ as
\begin{equation}
	\label{eq1:signal}
	S(x,y;z, z_d) = S_{SS}(x,y;z, z_d)+ S_{MS}(x,y;z, z_d), 
\end{equation}
where $x$ and $y$ are the lateral positions of the scanning locations, and $z_d$ is the amount of defocus at the imaging depth $z$. 
$S_{SS}$ and $S_{MS}$ are the signal components originating from SS and MS photons, respectively.
According to the formulation of Ralston \etal \cite{Ralston_2005}, the $S_{SS}$ component can be expressed as a convolution of a defocus-free OCT SS signal and a depth- and defocus-dependent quadratic phase function $\phi$ as
\begin{equation}
	\label{eq:defocusedSignal}
	S(x,y;z, z_d) = S_{SS}(x,y;z, 0) *  \exp \left[i \phi(x,y;z, z_d) \right]+ S_{MS}(x,y;z, z_d),
\end{equation}
where $S_{SS}(x,y;z, 0)$ is the defocus-free SS OCT signal.

\begin{figure} %[htbp]
	\centering\includegraphics[width=9cm]{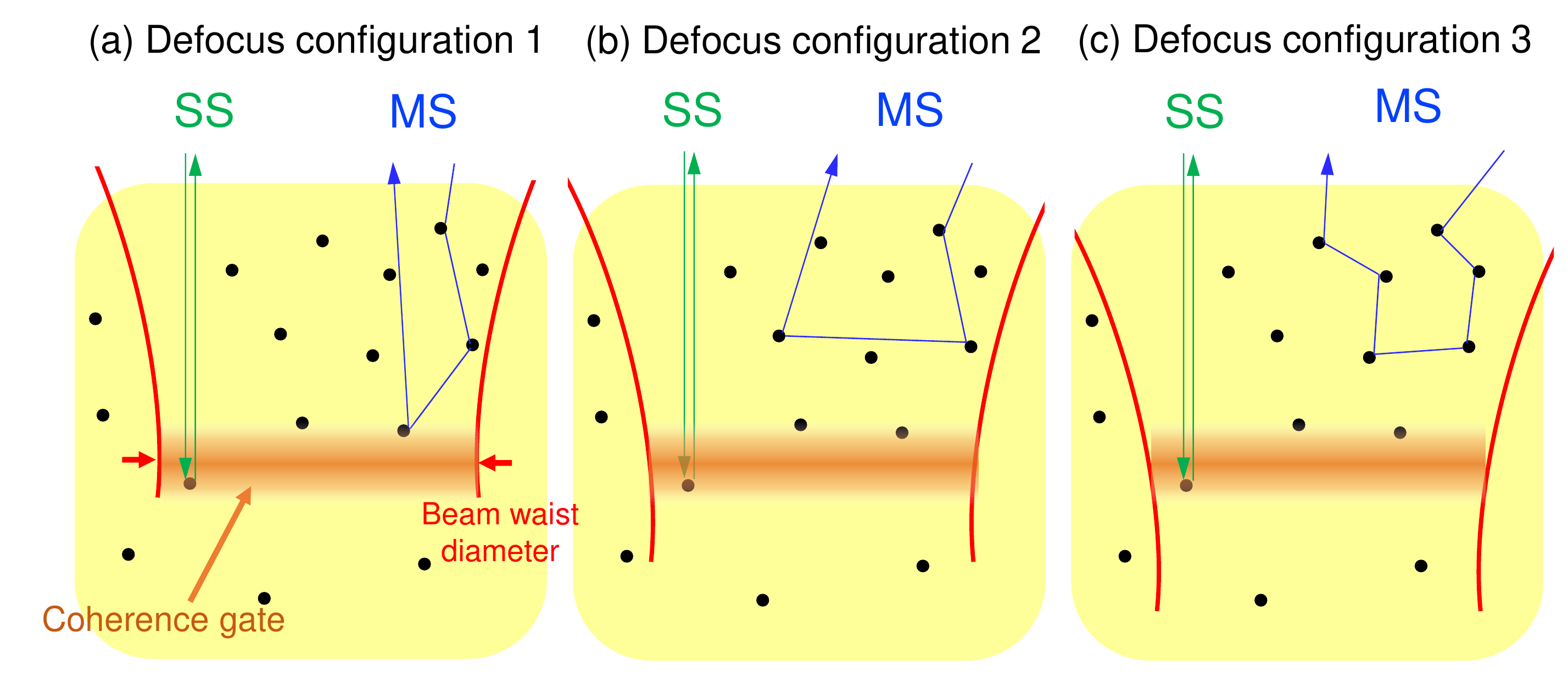}
	\caption{
		Schematics of trajectories of SS and MS photons with different focus depths; i.e., different defocus amounts.
		No matter the focal position, the SS photon is scattered only once and hence has the same path length.
		This results in the consistent phase of the SS signal after computational refocusing.
		Meanwhile, the trajectories of MS photons are scrambled by changes in the focal position. 
		Hence, the phase of the MS signal is randomized even after the computational refocusing.
	}
	\label{fig:principle}
\end{figure}
The SS photons experience only a single scattering event, and hence, they take consistent trajectories among the multiple measurements with different defocus, as depicted by the SS trajectories in Fig.\@ \ref{fig:principle}.
Hence, once the defocus is corrected by computational refocusing, the phase of SS components become consistent among the volumes.

Meanwhile, the MS photons can be scattered by different combinations of scatterers if the defocus changes the MS trajectories as shown in Fig.\@ \ref{fig:principle}.
Hence, the trajectories of the MS photons are altered by the defocus.
This alteration of the trajectories results in an inconsistent phase of $S_{MS}$ components for the different defocus, even after computational refocusing.
These properties of the signal components are formulated as follows.

Computational refocusing can be described as the deconvolution of the quadratic phase function $\phi$.
After the computational refocusing, the OCT signal becomes
\begin{equation}
	\label{eq:refocusedSignal}
	S'(x,y;z, z_d) = S_{SS}(x,y;z, 0) + S_{MS}(x,y;z, z_d) * \exp \left[-i \phi(x,y;z, z_d) \right],
\end{equation}
where the deconvolution is represented as the convolution with the complex conjugate of $\phi$ (see the Appendix for details of this representation).
No matter the original defocus amount, the computational refocusing provides the consistent defocus-free SS component.
In addition, the signal strength of the SS component is enhanced by the refocusing.
Meanwhile, the phase of the MS component depends on the original defocus amount because the MS trajectories depend on the defocus amount.
Furthermore, because the scattering events of the MS photons occur at shallower depths than the SS components, the MS signal may have a defocus amount different from that of the SS signal.
Hence, the computational refocusing tailored for the SS component does not enhance the signal strength of the MS component.

In our MFA method, an MS-signal-suppressed OCT volume is obtained by complex averaging the multiple volumes with different defocus amounts after the computational refocusing as 
\begin{equation}
	\label{eq:mfaSignal}
	\overline{S'}(x,y;z) = S_{SS}(x,y;z, 0) + \frac{1}{N} \sum_{j=0}^{N-1} S_{MS}(x,y;z, z_{d, j}) * \exp \left[-i \phi(x,y;z, z_{d,j}) \right],
\end{equation}
where $N$ is the number of volumes being averaged and $z_{d,j}$ is the defocus amount of the $j$-th volume.
As the phase of the MS term (the second term) is unpredictable and practically random, the amplitude of this term is reduced by the averaging.
Meanwhile, the phase of SS term (the first term) is consistent among the volumes, and hence, its amplitude is not reduced.

In practice, a larger difference in the defocus amounts among the volumes may cause higher mutual randomness (i.e., higher decorrelation) of the phases of the MS components, and a better suppression of MS components can thus be achieved.
This hypothesis is experimentally validated in Section \ref{sec:discuss:protocol}.

\subsection{Implementation of the MFA method}
\subsubsection{OCT setup}
\label{sec:system}
In this study, we used a polarization sensitive OCT (PS-OCT) system, which has been described in Refs.\@ \cite{li_BOEB_2017, miyazawa_BOEB_2019}. 
However, we only used a single-polarization-channel OCT image (i.e., a conventional non-polarization sensitive OCT image) to demonstrate the MFA method.
The light source was a sweeping laser source (AXP50124-8, Axsun Technologies, MA) with a central wavelength of 1.31-\um and a scanning rate of 50 kHz. 
The objective (LSM03, Thorlabs, NJ) had an effective focal length of 36 mm and gave a depth of focus (DOF) of 0.36 mm in air. 
The lateral and axial resolutions were respectively 18 \um and 14 \um in tissue. 

An electrical tunable lens (ETL, EL-10-30-CI-NIR-LD-MV, Optotune Switzerland AG, Switzerland) was set on the sample arm to modulate the focal position. 
It is noteworthy that the ETL can be easily aligned and applied to standard OCT systems. 

\subsubsection{Core methods of the MFA}
\label{sec:principle:implement}
In our MFA method, multiple complex OCT volumes are acquired with different focal positions. 
For each volume, the bulk phase error is estimated and corrected using a smart-integration-path \cite{oikawa_BOEB_2020} method. 

The defocus is corrected in each volume by applying computational refocusing, where a series of phase-only deconvolution filters are applied at all depths to correct the defocus. 
Details of the refocusing method can be found in Section 2.2 of Ref.\@ \cite{zhu_BOE_2022}. 

When modulating the focal position using an ETL, the deformation of the lens affects the optical path length of the probe beam. 
This introduces an axial image shift, which sometimes reaches a few tens of microns among the focus-modulated volumes. 
In our MFA method, the axial shift is estimated and corrected using a sub-pixel intensity cross-correlation method\cite{hillmann_OE_2012}. 
In this process, B-scans from different volumes at the same location are extracted, and four-fold up-sampling along the depth direction is conducted. 
Intensity-based cross-correlations among the extracted B-scans are performed to estimate shifts with resolution of 1/4 of a pixel along the axial direction. 
As this image shift is consistent within each volume, the estimated shifts are used for co-registration of all B-scans in each volume. 

The inter-volume phase offsets are also corrected. 
The offsets are computed for each A-line using an intensity-weighted phase difference averaging.
In this averaging, one volume is used as a reference, and the product of the reference and the complex conjugate of another volume is computed. 
The product is then averaged in complex form along the depth direction within a certain depth range, which was 30 pixels with sufficient intensity in the present implementation.
The phase of the averaged signal is the phase offset. 
The phase offsets are then computed for each A-line and corrected.

Finally, all volumes are averaged in complex form to obtain an ``MFA volume.''  
The OCT intensity volume is then generated from the MFA volume.

\subsection{Validation study design}
\subsubsection{Samples}
\label{sec:sample}
\begin{figure}
	\centering\includegraphics[width=10cm]{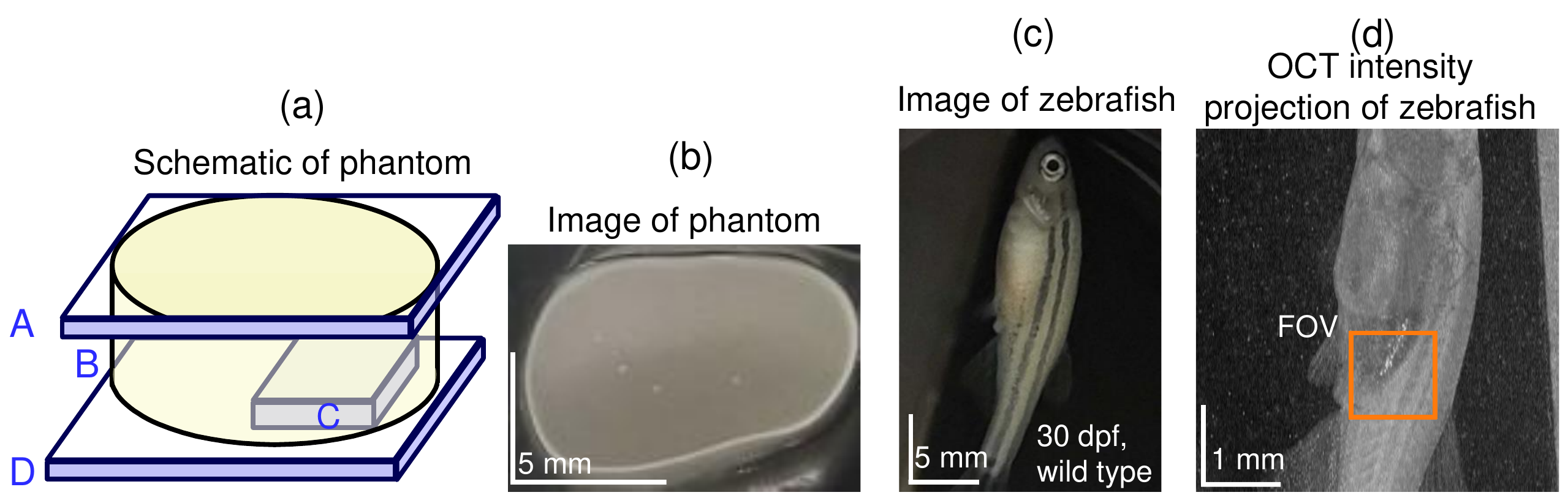}
	\caption{
		Schematic (a) and photograph (b) of a scattering phantom.
		The phantom comprises glass slips A and D, a scattering layer B, which is a mixture of polystyrene micro-particles and ultrasound gel, and a glass plate C embedded in the scattering layer.
		The glass plate C provides a scattering-free area.
		A postmortem zebrafish at 30 days post fertilization (dpf) was used as a biological sample.
		The sample is shown in a color photograph (c) and a wide-field-of-view OCT intensity projection (d).
		The orange box around the belly region denotes the measurement area in the validation study of the MFA method. 
	}
	\label{fig:samples}
\end{figure}
Two types of sample were measured to validate the performance of the MFA method. 
One was the scattering phantom illustrated in Fig.\@ \ref{fig:samples}(a), which comprises glass slips (A, D), a scattering layer (B), and a glass plate buried at the bottom of the scattering layer (C). 
The glass slip at the top (A) was tilted to prevent a strong reflection from the surface. 
The scattering layer (B) was a mixture of 0.025 mL of 10\%-concentration polystyrene micro-particles (diameter of 10 \um, 72968-10ML-F, Sigma-Aldrich) and 0.6 mL of ultrasound gel (Pro Jelly, Jex, Japan). 
The glass plate (C) had a thickness of approximately 0.12 mm and was buried at the bottom of the scattering layer to create a space without scattering. 
The thickness of the scattering layer above the glass plate was approximately 1.5 mm. 
A photograph of the sample from the top is shown in Fig.\@ \ref{fig:samples}(b).

A 30-dpf wild-type postmortem female zebrafish [Fig.\@ \ref{fig:samples}(c)] was imaged to demonstrate the measurement of a biological sample.
The zebrafish was anesthetized by tricaine and sacrificed by low-temperature treatment (placed in ice for 2 minutes) and then placed in a petri dish and immersed in saline solution for measurement. 
A piece of black tape was placed on the bottom of the petri dish to prevent strong specular reflection.
A wide-field-of-view \enface OCT-intensity projection of the zebrafish is shown in Fig.\@ \ref{fig:samples}(d), where the orange box indicates the measured area. 

The postmortem zebrafish measurement was performed following the animal experiment guidelines of the University of Tsukuba. 
The measurement protocol was approved by the Institutional Animal Care and Use Committee (IACUC) of University of Tsukuba. 

\subsubsection{Measurement protocol}
\label{sec:measureprotocl}
In both the phantom and zebrafish measurements, the samples were placed on a linear translation stage and measured. 
Seven volumes with different focal positions were acquired for averaging. 
The interval between focal positions is referred to as the focus shifting step $\deltaz$ and was set at 0.12 mm (approximately $1/3$ of the DOF).
The overall shifting distance of the focus is denoted as $D = (N-1) \deltaz$ and was 0.72 mm in the study. 
These parameters were determined empirically, and details of the parameter selection strategy are given in Section \ref{sec:discuss:protocol}. 
For comparison, another set of raw volumes were acquired and averaged without focus modulation.
This averaged volume without focus modulation is referred as the ``complex averaging'' volume . 

In the measurements, the focal positions were set in the deep regions of the samples to enhance the light collection efficiency.
The lateral scanning range was 1.5 mm $\times$ 1.5 mm sampled with 512 $\times$ 512 A-lines, which gave an isotropic lateral pixel separation of 2.93 \um (around 1/6 of the lateral optical resolution). 
The acquisition time for each volume was approximately 6.5 s. 

\subsubsection{Quantitative image contrast analysis}
We defined the SBR to quantify the image contrast in the scattering phantom measurement. 
An \enface image at a depth slightly beneath the top surface of the buried glass plate was extracted. 
Several scatterers with similar intensities were manually selected to compute the SBR.
The signal intensity of each scatterer was defined as the averaged image intensity within a 3-pixel $\times$ 3-pixel window centered at the scatterer.
The signal intensity used to compute the SBR was then defined as the mean of the signal intensities of the selected scatterers.
The background intensity was defined as the mean intensity of pixels in the buried glass plate region (i.e., the scattering-free region).

\section{Results}
\label{sec:result}
\subsection{Scattering phantom}  
\label{sec:result:phantom}
\begin{figure}
	\centering\includegraphics[width=12cm]{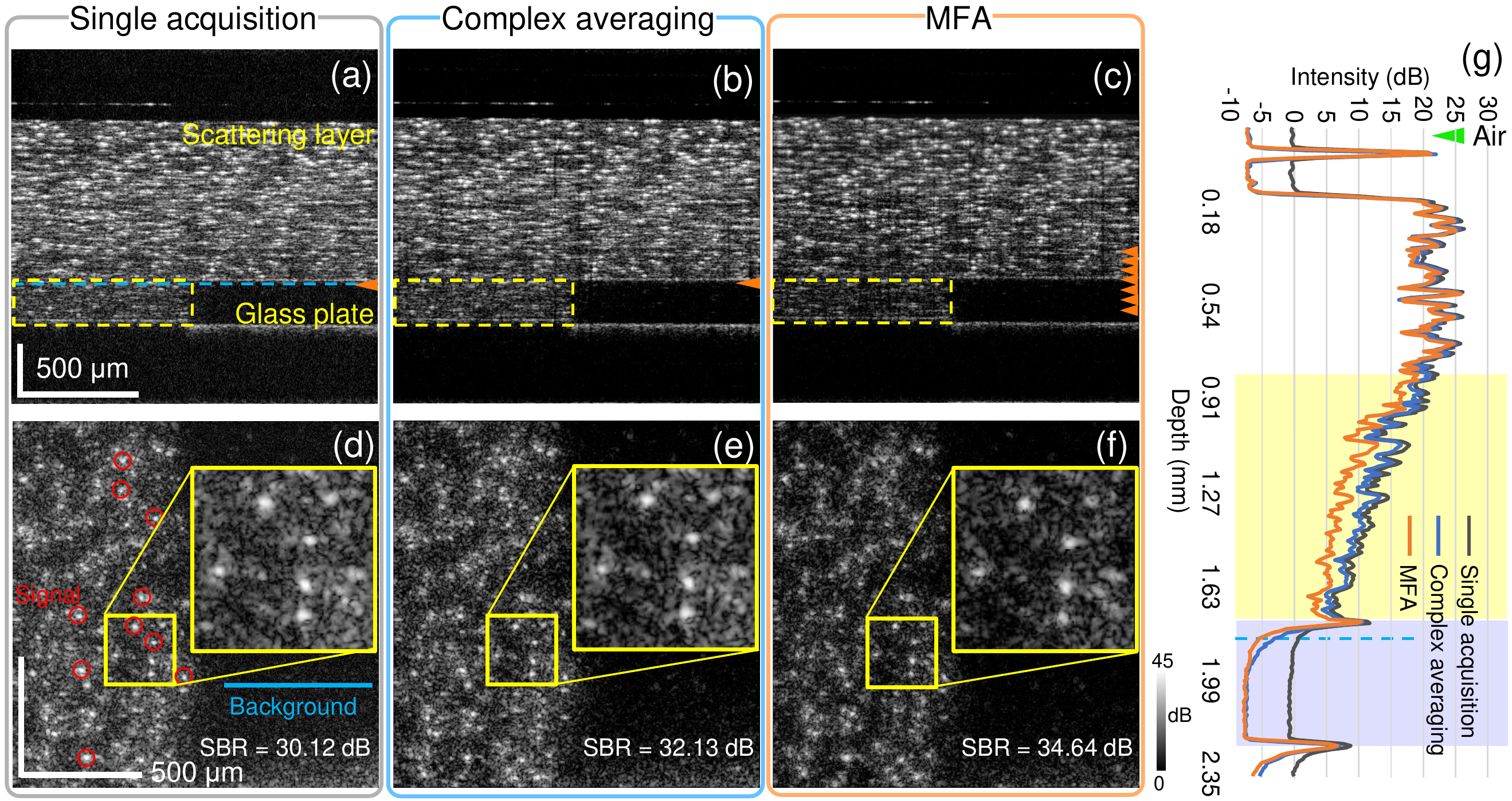}
	\caption{
		(a)--(c) and (d)--(f) show the B-scans and \enface images of the phantom, respectively. 
		In (a)--(c), the orange arrowheads denote the approximate focal positions, and the yellow boxes denote a deep region of the scattering layer. 
		In (d)--(f), the yellow boxes and the magnification insets are the \enface images of a small region in the scattering layer containing several scatterers. 
		In (d), the red circles denote the scatterers selected to compute the signal intensity.
		The blue line indicates a region with 2 $\times$ 200 A-lines in the glass plate region used to compute the background intensity for the SBR analysis. 
		(g) shows the intensity depth profiles that are averaged by the A-lines at the \enface locations denoted by the blue line in (d). 
		The 0-mm depth refers to the top surface of the cover glass on the top, and the green arrowhead denotes the depth in air where the intensity is 0 dB. 
		The blue dashed lines in (a) and (g) denote the depth where the \enface images are taken. 
	}
	\label{fig:phnatom}
\end{figure}

Figure \ref{fig:phnatom}(a)--(c) shows the OCT intensity B-scans of the single acquisition, complex averaging, and MFA methods, respectively. 
The approximate focal positions are denoted by the orange arrowheads. 
Multiple orange arrowheads in Fig.\@ \ref{fig:phnatom}(c) denote different focal positions of each image used to generate the MFA image.  
Among the B-scans, the MFA image provides a lower background noise in the scattering layer than the other images, especially in the deep regions (denoted by the yellow dashed boxes). 

Figure \ref{fig:phnatom}(g) presents the intensity depth profiles obtained using the three methods. 
The depth intensity profile of the MFA (orange) shows a lower intensity than the other profiles as depth increasing (the region with yellow background).
Since the MS signal is expected to be more pronounced in deeper regions, this profile may indicate the MS suppression by MFA method.
Meanwhile, the single acquisition (black) and complex averaging (blue) profiles have intensities similar to each other, which indicates that the complex averaging method does not appreciably reduce MS. 
At the depths of the buried glass plate region (blue background region) in Fig.\@ \ref{fig:phnatom}(g), we see that the MFA and complex averaging methods show appreciably lower signal intensity than that of the single acquisition method. 
It may be because of the suppression of measurement noise by the averaging. 
At a depth close to the top surface of the buried glass plate (indicated by a dashed blue line), the complex averaging curve has a lifted ``tail'' with intensity higher than that in the air.
As the glass plate region is scattering-free, this tail might be due not to scattering in the glass but to the MS in the superior scattering part of the phantom.
A similar result was reported by Yadlowsky \etal \cite{yadlowsky_AOA_1995}.
This lifted tail is reduced when using the MFA method, which further supports the MS-suppression ability of MFA.
These findings suggest that the complex averaging method reduces the system noise, and the MFA method reduces both system noise and MS signals. 

Figure \ref{fig:phnatom}(d)--(f) show the \enface images at the same depth indicated by the blue dashed lines in Figs.\@ \ref{fig:phnatom}(a) and (g).
%The scatterers appear with a better contrast in the MFA image than in the others (see the magnified images within the yellow boxes). 
The complex averaging image has slightly better contrast of the scatterers than the single acquisition image, but this contrast is not as good as that of MFA image (magnifications in the yellow boxes). 
The MFA method provides better image contrast than the complex averaging method because the former benefits from the MS reduction.

The SBR is computed from Fig. \ref{fig:phnatom}(d)--(f) for quantitative comparison. 
Ten scatterers are selected to compute the signal intensity [red circles in Fig. \@\ref{fig:phnatom}(d)]. 
The background intensity is computed from the pixels in the glass plate region [blue line in Fig. \@\ref{fig:phnatom}(d)]. 
The computed SBRs are 30.12, 32.13, and 34.64 dB for single acquisition, complex averaging, and MFA images, respectively. 
Adopting the current protocol, the MFA method provides an SBR improvement of 4.5 dB relative to the single acquisition, and 2.5 dB relative to the complex averaging method.

\subsection{Postmortem zebrafish}  
\label{sec:result:zebrafish}
\begin{figure}
	\centering\includegraphics[width=13cm]{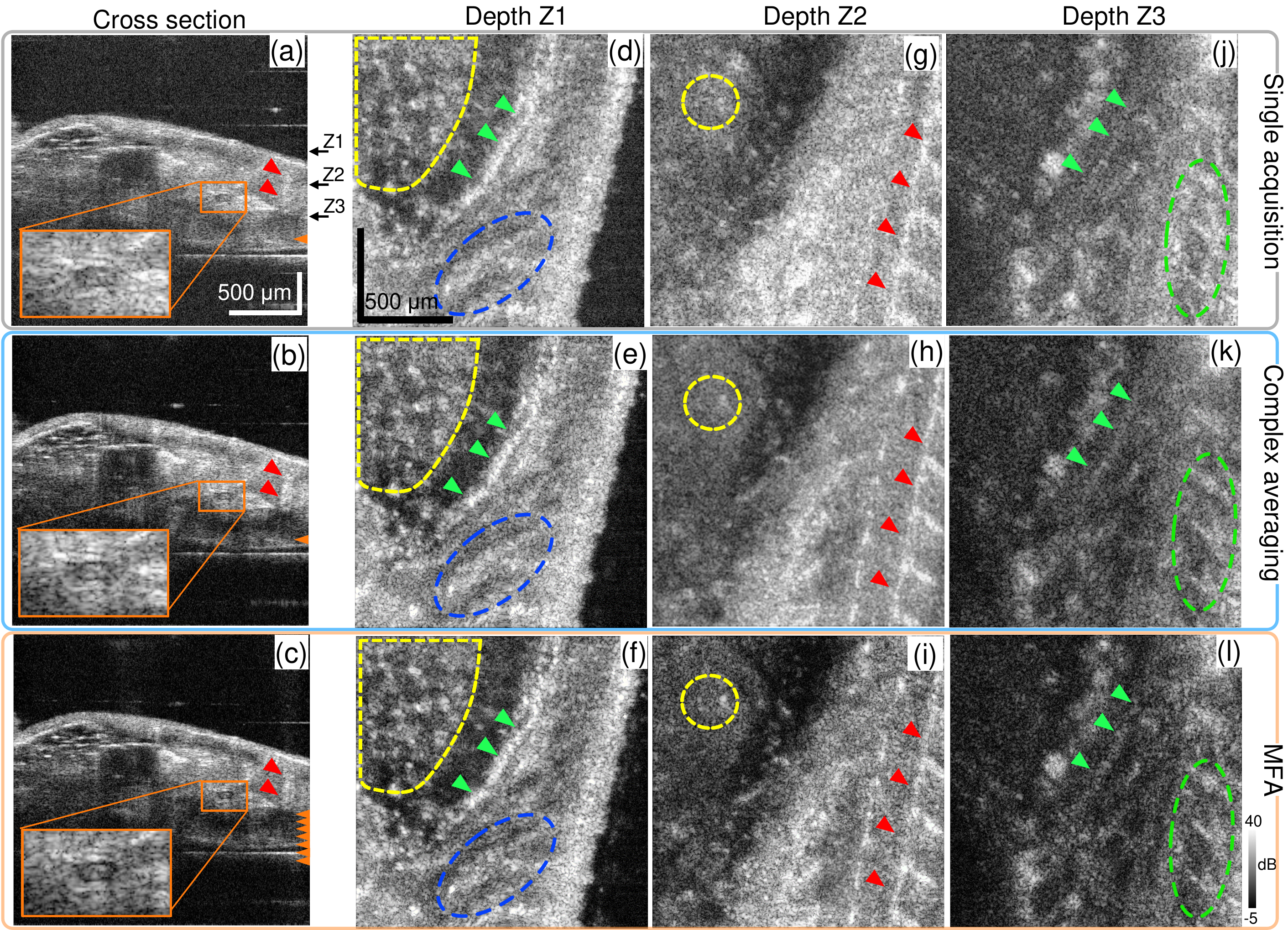}
	\caption{
		Cross sectional and \enface images of the postmortem zebrafish. 
		In (a)--(c), the orange arrowheads denote the approximate focal positions, and the orange boxes and magnification insets are supposed to show notochord structure. 
		The red arrowheads in (a)--(c) and (g)--(i) may indicate the myosepta. 
		In (d)--(f), yellow dashed areas denote the belly region, and blue ellipses mark fine structures observed in the muscle region. 
		The green arrowheads in (d)--(f) and (j)--(l) may indicate the outer layer of the swim bladder. 
		In (g)--(i), yellow circles denote several hyper-scattering spots observed in the belly region. 
		In (j)--(l), green ellipses denote unidentified structures that are better contrasted in the MFA image. 
		At all depths, the MFA method has the images with best contrast among the three methods. 
	}
	\label{fig:fish}
\end{figure}
Figure \ref{fig:fish}(a)--(c) shows the B-scans of the sample. 
The MFA images better visualize structural boundaries [orange magnification inset in Fig.\@ \@\ref{fig:fish}(c)], which are smeared by strong noise in Figs.\@ \ref{fig:fish}(a) and (b). 
This low-scattering-intensity structure is expected to be the notochord according to its anatomic features because a similar transparent structure of the notochord has been observed at 5 dpf\cite{Haindl_2020}.  

Figure \ref{fig:fish}(d)--(f) shows the \enface images at depth Z1. 
The MFA image provides better contrast for some fine structures in the muscle region (within blue dashed ellipses) and some hyper-scattering spots in the belly region (yellow dashed areas). 

At depth Z2 [Fig.\@ \ref{fig:fish}(g)--(i)], the contrast between the myosepta (red arrowheads) and the surrounding muscle tissues is again highest for the MFA method among the three methods.
Some hyper-scattering spots in the belly region (yellow dashed circles) are sharp and recognizable in the MFA image, whereas they are blurred in the complex averaging image.
These hyper-scattering spots are almost unrecognizable in the single acquisition image.

At a deeper depth Z3, a thin hyper-scattering tissue is noted in the MFA image [green arrowheads in Fig.\@ \ref{fig:fish}(l)] but difficult to recognize in the single acquisition and the complex averaging images [Figs.\@ \ref{fig:fish}(j) and (k)]. 
This tissue is also observed at the upper depth Z1 [green arrowheads in Fig.\@ \ref{fig:fish}(d)--(f)]. 
A similar structure has been visualized in adult zebrafish by OCT\cite{Yang_BOE2020}, and it is expected to be the outer layer of the swim bladder. 
Unidentified structures elongated along the dorsal side of the zebrafish [green ellipses in Fig.\@ \ref{fig:fish}(j)--(l)] are also visualized with the highest contrast in the MFA image. 

These findings suggest that the MFA method is applicable to biological samples, reduces the MS signal, and enables better visualization of structures.

%%%%%%%%%%%%%%%%%%%%%%%%%%%%%%%%%%%%%%%
\section{Discussions}
\subsection{Validation of the induced defocus and its correction by MFA method}
\label{sec:discuss:resolution}
\begin{figure}
	\centering\includegraphics[width=13cm]{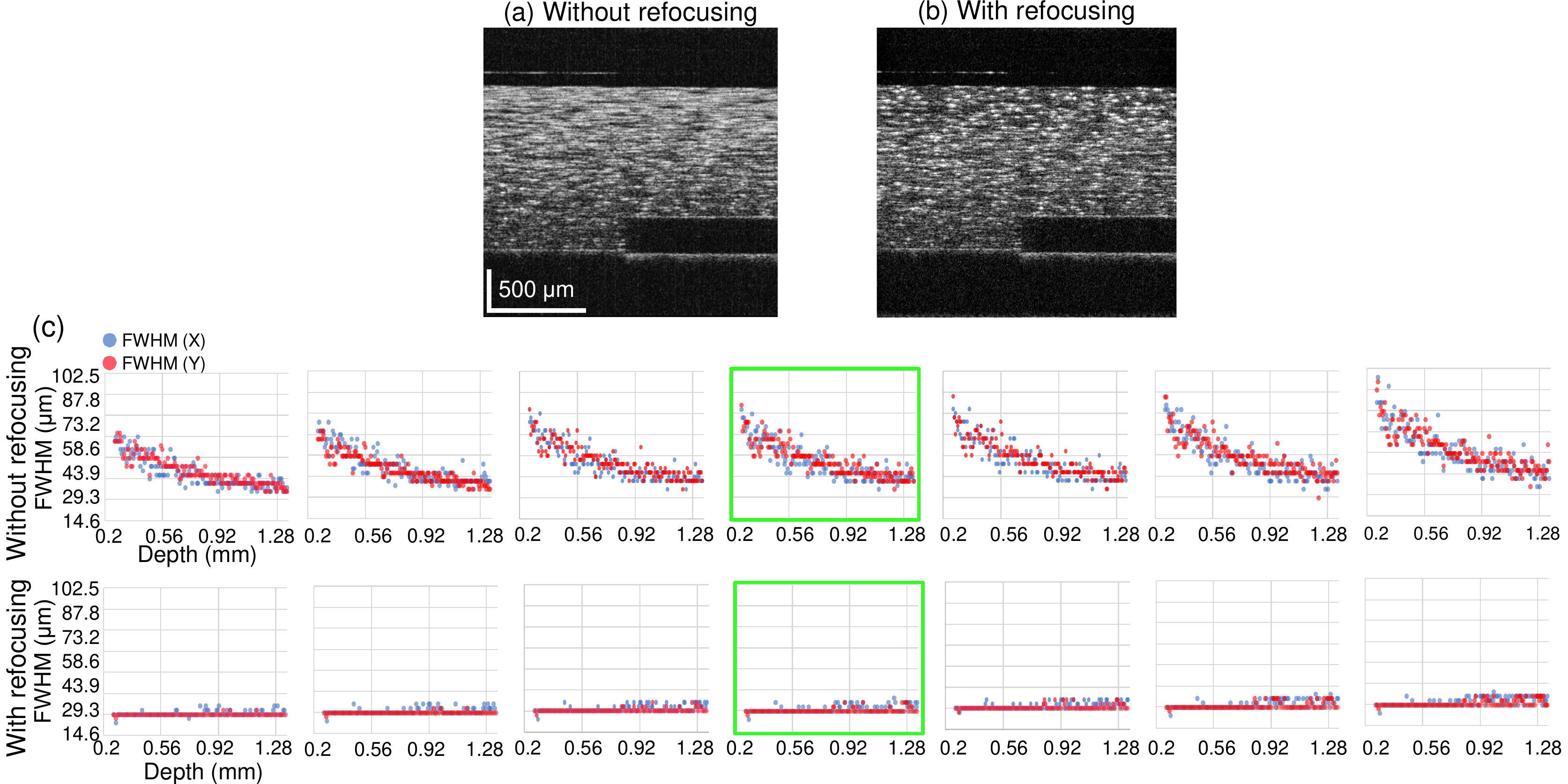}
	\caption{
		(a) and (b) are B-scans at the same location of a phantom volume without and with refocusing, respectively. 
		(c) shows the FWHMs of the spatial autocorrelation functions of the linear-scaled images, which are considered to be proportional to the speckle size.
		The first and second rows present the results without and with computational refocusing, respectively.
		Each column presents the results for a volume measured with different focal positions.
		The blue and red plots show the FWHMs for the fast-scanning (X) and slow-scanning (Y) directions.
		The B-scans (a) and (b) were taken from the volumes corresponding to the plots highlighted by green boxes.
		The used volumes are identical to those used for the MFA images in Fig.\@ \ref{fig:phnatom}. 
	}
	\label{fig_resolution}
\end{figure}
To validate the defocus correction over multiple acquisitions, we compared the lateral resolutions of a set of seven OCT volumes with different focal positions.
These volumes are the same phantom volumes used for the MFA method demonstration in Section \ref{sec:result:phantom} (Fig.\@ \ref{fig:phnatom}).
Figures \ref{fig_resolution}(a) and (b) show representative B-scans of the phantom without and with refocusing, respectively. 
Note that Fig.\@ \ref{fig_resolution}(b) is identical to Fig.\@ \ref{fig:phnatom}(a). 

The lateral resolution at each depth was evaluated by the speckle size of the \enface intensity image, which was estimated using a linear-intensity-based auto-correlation.
The speckle size was defined as the full width at half-maximum (FWHM) of the auto-correlation function.
This estimation was performed along both fast-scanning (denoted as ``X'') and slow-scanning (denoted as ``Y'') directions.
A depth range of 1.1 mm (150 pixels) beneath the surface of the scattering layer was used for this validation. 

Without refocusing, the seven volumes with different defocus have speckle sizes (X and Y FWHMs) varying along the depth [Fig.\@ \ref{fig_resolution}(c), first row].
After applying refocusing to each volume, the FWHMs become almost constant at each depth [Fig.\@ \ref{fig_resolution}(c), second row].
Note that the horizontal lines in the plot are not regression or theoretical curves but plotted dots appearing as lines because the FWHMs are highly constant.
Most of the FWHMs converge on a value of 14.6 \um, whereas the system optical lateral resolution (defined as the focus radius where the intensity falls to $1/e^2$ of the maximum) is 18 \um. 
This suggests that computational refocusing corrects the defocus over a depth range exceeding 1 mm, regardless of how much defocus is applied in the measurement.

\subsection{Focus shifting protocol optimization}
\label{sec:discuss:protocol}
\subsubsection{Experimental optimization of MFA parameters}
\label{sec:discuss:protocol:optimize}

\begin{table}
	\caption{
		Summary of parameters used in the parameter-optimization experiment. 
		Parameters $\deltaz$, $N$, and $D$ are the focus shifting step, number of averaged volumes, and overall focus shifting distance, respectively. 
		The DOF corresponds to 0.36 mm of the axial distance in air. 
	}
	\centering\scalebox{0.7}{
		\begin{tabular}{ccll} \hline
			Measurement configuration & $\deltaz$ [$\times$ DOF] & $N$          & $D=(N-1) \deltaz$ [$\times$ DOF]    \\  \hline
			\#1 &  1         & 1, 2, 3, 4, 5    & 0, 1, 2, 3, 4  \\
			\#2 &  1/2    & 1, 2, 3, ..., 5, ..., 9  & 0, 0.5, 1, ..., 2, ..., 4   \\
			\#3 & 1/3     & 1, 2, 3,..., 13 & 0, 0.33, 0.36,..., 4    \\
			\#4 & 1/4     & 1, 2, 3,..., 13 & 0, 0.25, 0.5,..., 3    \\
			\#5 &  1/6     & 1, 2, 3,..., 13 & 0, 0.17, 0.33,..., 2   \\ \hline         
	\end{tabular} }
	\label{tab1:protocol}
\end{table}

\begin{figure}
	\centering\includegraphics[width=10cm]{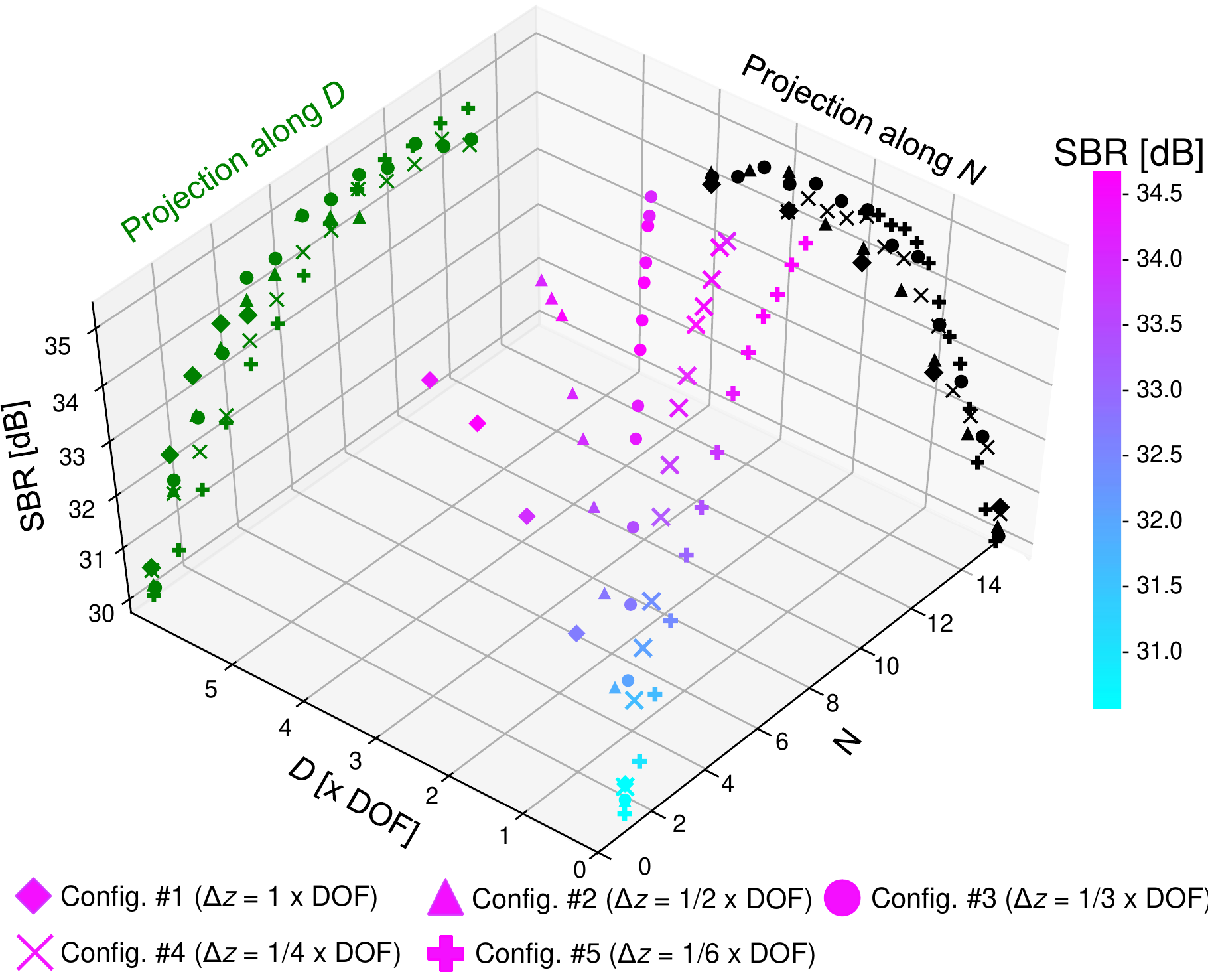}
	\caption{
		SBR plot against the number of averaged volumes $N$ and the overall focus shifting distance $D$.
		The type of plotted dot indicates the focus shifting step $\deltaz$.
		The plot was produced to select optimal parameters of $N$, $D$, and $\deltaz$. 
		The measured sample is the phantom shown in Fig.\@ \ref{fig:phnatom}, and the SBR was computed by the process described in Section \ref{sec:result:phantom}. 
		The black and green plots are projections showing the relationships between the SBR and $D$ and between the SBR and $N$, respectively.
	}
	\label{fig_sbr}
\end{figure}
To find the optimal protocol, we experimentally explored the relationship between the number of averaged volumes $N$, the overall focus shifting distance $D$, the focus shifting step $\deltaz$, and the SBR. 
In this experiment, we performed five sets of sequential volume measurements with different $\deltaz$ as summarized in Table \ref{tab1:protocol}.
For each set of measurements, we extracted subsets to examine several configurations of $N$. 
$D$ is determined from  $\deltaz$ and $N$ as $D=(N-1)\deltaz$.
The sample used for this measurement was the phantom shown in Fig.\@ \ref{fig:phnatom}.

As there are two independent parameters, we plot the SBRs in a three-dimensional space of the SBR, $D$, and $N$ as shown in Fig.\@ \ref{fig_sbr}.
The color gradient of the plotted points represents the SBR.
The SBR takes a minimum value when $N$ and $D$ are close to zero and increases with both $N$ and $D$. 
The black plot is the projection along the $N$-axis that shows the relationship between the SBR and $D$. 
We see that the SBR saturates at $D$ approximately twice the DOF and starts to drop at $D$ exceeding approximately 4 times the DOF. 
The green plot shows another projection along the $D$-axis that gives the relationship between the SBR and $N$. 
It is seen that the datasets with different $\deltaz$ saturate at different $N$.
For example, the datasets of configuration \#1 ($\deltaz = 1 \times \mathrm{DOF}$) and configuration \#4 ($\deltaz = 1/4 \times \mathrm{DOF}$) saturate at approximately $N$ = 4 and $N$ = 8, respectively. 
These results suggest that both the parameters $N$ and $D$ contribute to the SBR improvement in the MFA method. 

\begin{figure}
	\centering\includegraphics[width=13cm]{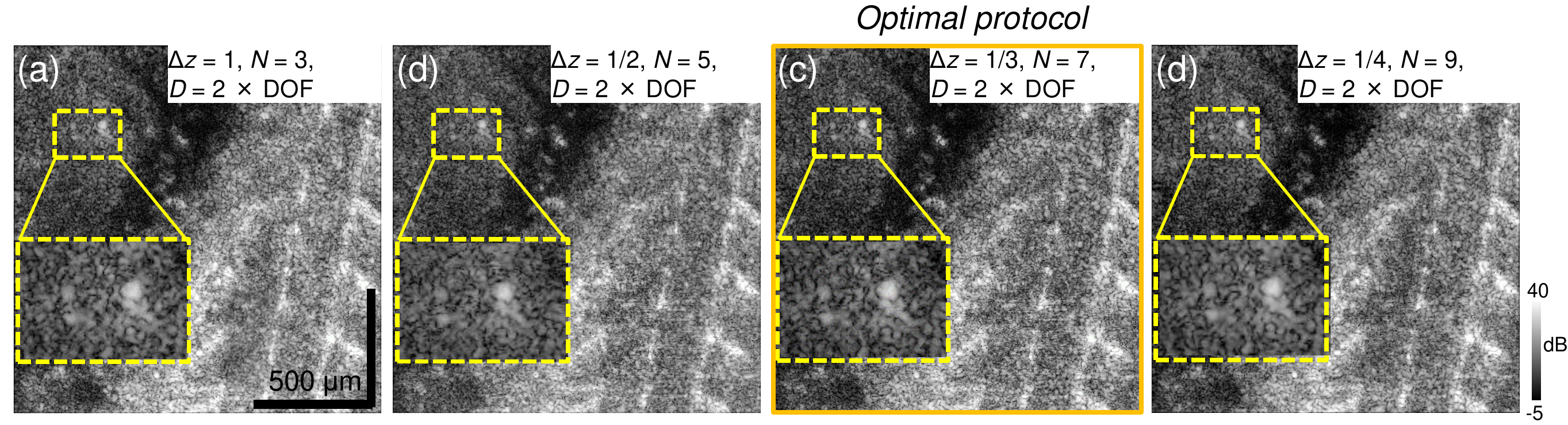}
	\caption{
		\Enface images of the zebrafish measured with different MFA parameters at the same depth and the same lateral position.
		Among the four sets of parameter configurations, configuration (c) gives best contrast.
		This configuration was thus selected as the optimal parameter set in the present study. 		
	}
	\label{fig_fish_protocol}
\end{figure}
The optimal parameters were selected based on the above results and used in the experiments described in the previous sections. 
We first selected $D$ as twice the DOF, where the projection curve starts saturating (black plots in Fig.\@ \ref{fig_sbr}).
We then performed measurements of the zebrafish sample with $\deltaz$ of 1, 1/2, 1/3, and 1/4.
As $D$ was fixed to be twice the DOF, $N$ was 3, 5, 7, and 9, respectively.
The imaging results (Fig.\@ \ref{fig_fish_protocol}) show that the parameters correpond to $\deltaz=1/3$, $D=2$, $N=7$ (third column) seems to provide best contrast of the hyper-scattering spots in the belly region (see the magnified insets within yellow boxes). 
We therefore selected this parameter set as our optimal protocol and used it to perform the measurements presented in the results section.  

\subsubsection{Interpretation of the SBR drop}
\label{sec:discuss:protocol:interpret}
In the protocol optimization, we note that the SBR drops when $D$ becomes too large.
A possible reason of this SBR drop is the uncorrected confocality; i.e., the signal amplitude drops when the sample plane is far from the focal plane. 
When $D$ exceeds a few DOFs, the focus shifts far from the depth of interest. 
As a result, the amplitudes of both SS and MS signals decrease even with the computational refocusing, whereas the amplitude of the measurement noise is unchanged regardless of the focal position. 
The solution to overcome this limit remains an open issue and requires further investigation.

%%%%%%%%%%%%%%%%%%%%%%%%%%%%%%%%%%%%%%%
\subsection{Advantages of electrical tunable lens (ETL)}
\label{sec:discuss:etl} 
\begin{table}
	\caption{
		Comparison of examples of wavefront manipulation devices.
		Although the ETL controls only defocus, it is the cheapest among the devices.
	}
	\centering\scalebox{0.67}{
		\begin{tabular}{lllll} \hline
			Device  &  ETL    & Deformable mirror  &  SLM  &  Deformable membrane  \\  \hline
			Type  & EL-10-30-CI-NIR-LD-MV    &   DM97-15   &   PLUTO NIR2  &  uDM2   \\  
			Manufacturer  & Optotune Switzerland  & Alpao France    &   HoloEye Photonics Germany  &  Dyoptyka Ireland  \\
			Manipulable order of wavefront & Defocus only & Till 4-th order & Very high order & Not applicable \\ 
			Response time  &   15 ms   &   1 ms (open-loop)   &   $\sim$ 30 ms  &  <1 ms   \\
			Approximate price  & \$ 600 &  \$ 30,000  &   \$ 10,000  &  \$ 3,500   \\ \hline         
	\end{tabular} }
	\label{tab2:DeviceCompare}
\end{table}

\begin{figure}
	\centering\includegraphics[width=11cm]{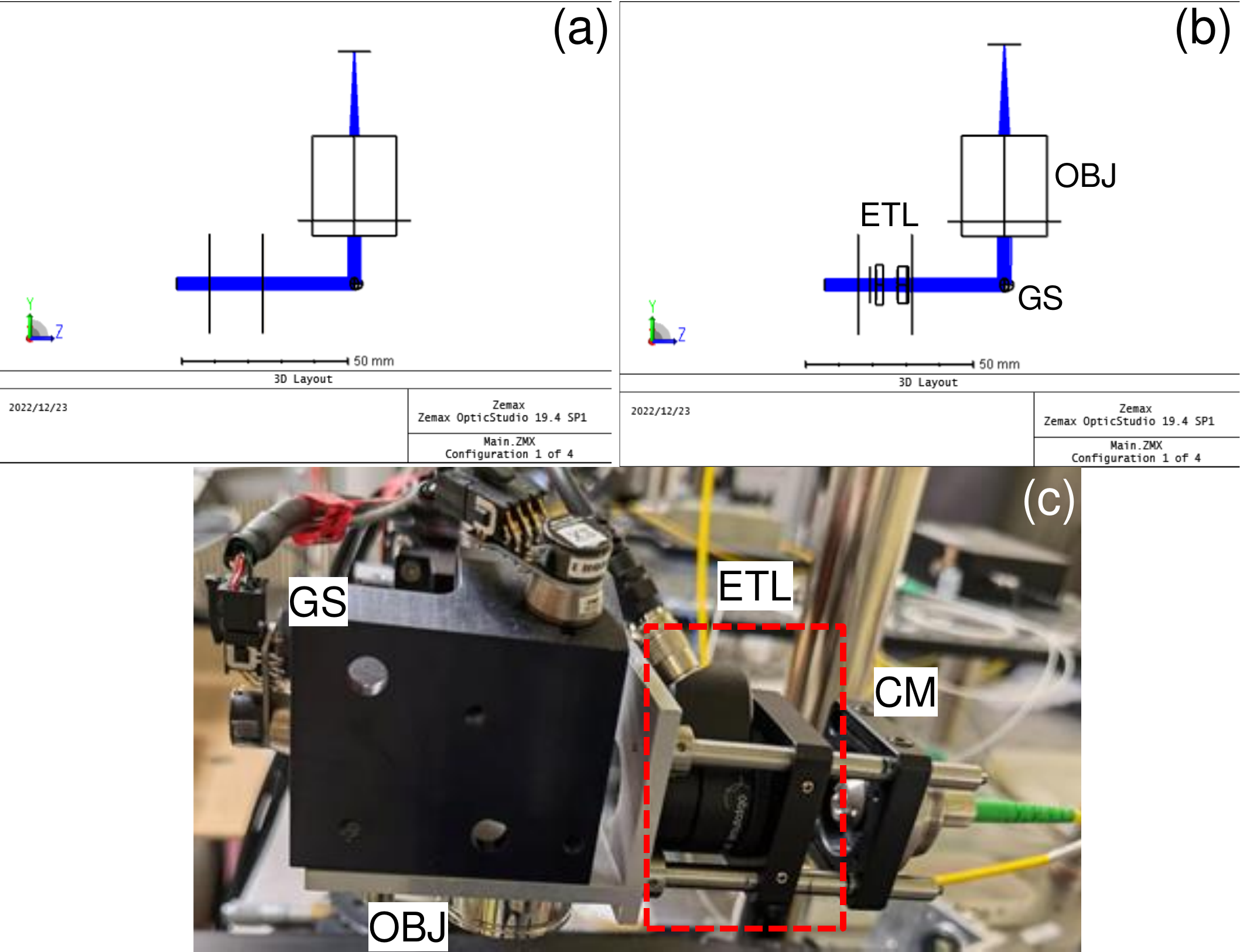}
	\caption{
		Optical design schematics of the sample arm without (a) and with (b) the ETL. 
		(c) Color photograph of the sample arm with the ETL. 
		GS: galvanometric scanner, OBJ: objective, ETL: electrical tunable lens, and CM: collimator.
		Since the ETL is a refractive optical element and easy to align, switching between the two configurations without and with the ETL takes only a few minutes. 		
	}
	\label{fig_schematic}
\end{figure}
Other wavefront manipulation devices have recently been used to reduce MS for improving the OCT image contrast.
One such device is a deformable mirror, which was used by Liu \etal in aberration-diverse OCT.
Here, the astigmatism was introduced to the illumination beam, and its astigmatic angle was modulated to decorrelate the MS among the measurements \cite{liu_BOEB_2018a}. 
In addition, a deformable membrane mirror has been combined with a full-field SS-OCT, and fast three-dimensional volumetric cross-talk-free imaging has been demonstrated \cite{auksorius_BOEB_2019, Stremplewski_Optica2019}.
Borycki \etal used a spatial phase modulator for digital aberration correction and MS reduction \cite{borycki_OLO_2020}. 

An advantage of the ETL over these wavefront manipulation devices is the low cost of the ETL. 
A simple comparison of the aforementioned devices is shown in Table \ref{tab2:DeviceCompare}. 
Another advantage of the ETL is that the ETL is easily integrated with standard OCT systems because it is a refractive (not reflective) device. 
The optical design schematics of the sample arm without and with the mounting of the ETL in our system are shown in Fig.\@ \ref{fig_schematic}(a) and (b), respectively. 
Figure \ref{fig_schematic}(c) is a photograph of our sample arm, where the ETL (shown by red dashed box) was inserted between the collimator and the mount of the galvanometric scanner.
In our implementation, the replacement of the ETL and the realignment of the system usually takes only a few minutes. 
This simple implementation of the ETL makes our probe arm switchable between the standard OCT mode and MFA-compatible mode.

\subsection{Implementation in PS-OCT}
\label{sec:discuss:psoct}
PS-OCT is a functional extension of OCT that provides additional polarization contrast for the quantitative evaluation of the optical properties of tissues \cite{deBoer_BOE2017, Baumann_2017, Yasuno2023JSTQE}.
It is also known that MS affects the polarization measurement of PS-OCT \cite{Baumann_2017}. 
Adie \etal reported that the presence of MS randomized the polarization states\cite{adie_OEO_2007}, leading to so-called depolarization. 
Several researchers suspected that MS generates artifacts in accumulative or local phase retardation measurements in the imaging of biological samples \cite{Matcher_2004, Lichtenegger_2022}. 

However, the MS issue in PS-OCT has rarely been addressed. 
One example was demonstrated by Gao \etal in FF-OCT, who used a Muller-matrix-based MS subtraction process to remove the MS-induced local phase retardation\cite{Gao_JInnovOptHealthSci_2020}.

The MFA method is applicable also to PS-OCT.
Our preliminary study showed that in a scattering phantom, the measured degree-of-polarization uniformity (DOPU) decreases in deep regions, which can be mitigated by adopting the MFA method\cite{Zhu_spie2023}.  
We are currently working on adapting the MFA method to PS-OCT imaging.
This adaption may improve the accuracy of quantitative polarization measurements in biological tissues.

\subsection{\Invivo imaging}
\label{sec:discuss:invivo}
The ETL enables fast and accurate focus modulation with a mechanical response time of 15 ms \cite{optotune}. 
However, the current MFA method captures multiple volumes for averaging, which is time consuming and a challenging task for \invivo imaging. 

Nevertheless, the long acquisition time can be overcome by several means. 
One possible solution is to use ultra-high-speed acquisition OCT\cite{Klein2013BOE, Kumar_BOE2017, Kolb_PO2019} and a motion tracking module to reduce the effect of sample motion. % Tsai_BOE2013: VECSEL, 1Mhz
Another possible solution is to use B-scan-based focus modulation instead of modulating the focus for each volume.
In other words, we can capture multiple B-scans at the same location with different focal positions, applying a one-dimensional version of computational refocusing \cite{yasuno_OE_2006}, and average the focus-corrected B-scans to generate an MS-reduced B-scan. 
Finally, the three-dimensional volume can be obtained by repeating this process for a volume.
In this case, the acquisition time for generating an MS-reduced B-scan can be reduced if compared with the current MFA method. 
A proof-of-concept experiment showed that this method achieves a level of MS suppression similar to that achieved by the current MFA method \cite{Yiqiang_bios2023}.

\section{Conclusion}
We have developed a new method termed ``MFA'', to suppress the MS signals and improve the image contrast in OCT. 
A scattering phantom was measured to validate the MS reduction of the MFA method.
The contrast improvement of the MFA image was compared with single acquisition and complex averaging images quantitatively.
A postmortem zebrafish was measured to demonstrate the ability of the MFA method in biological imaging. 
We expect this proposed method will help better visualize the anatomic features. 
MFA method has great potential of being adopted in \invivo imaging and other imaging modalities to reduce the multiple-scattering effect.  

\appendix
\section*{Appendix}
\section{Deconvolution of the quadratic phase function}
In Eq.\@ (\ref{eq:mfaSignal}), the deconvolution is represented as a convolution with the complex conjugate of the quadratic phase function of the defocus amount, $\phi$.
This representation is adopted because the Fourier transform of the quadratic phase function can be approximated by a phase-only function \cite{yasuno_OE_2006} and $\phi$ is an even function.
In this case, the deconvolution is expressed by the convolution of the complex conjugate of $\phi$ as follows.

In general, the convolution of two functions $f(x)$ and $g(x)$ is expressed as a multiplication in Fourier space as 
\begin{equation}
	\ftcal{f(x) * g(x)} = F(\nu)  G(\nu),
\end{equation}
where $F(\nu)$ and $G(\nu)$ are the Fourier transforms of $f(x)$ and $g(x)$, respectively.
$\nu$ is the Fourier pair of $x$, and $x$ is not a specific variable of space but a general variable.

Meanwhile, the deconvolution of $g(x)$ from $f(x)$ is expressed as the multiplication $F(\nu) G^{-1}(\nu)$ in the Fourier space.
When $G(\nu)$ is a phase-only function, as in the case of the Fourier transform of the quadratic phase function $\phi$, $G^{-1}(\nu) = G^*(\nu)$, where the superscript $*$ denotes the complex conjugate.
Hence, the deconvolution of $g(x)$ from $f(x)$ can be written as the convolution $f(x) * g^*(-x)$, where we use a general property of the Fourier transform, namely the inverse Fourier transform of $G^*(\nu)$ is $g^*(-x)$.

When $g(x)$ is an even function, as in the case of the quadratic phase function $\phi$, the deconvolution by $g(x)$ can be expressed as the convolution with $g^*(x)$ as $f(x) * g^*(x)$.
Hence, the refocusing can be expressed as in Eq.\@ (\ref{eq:refocusedSignal}).

\section*{Acknowledgments}
Lida Zhu is supported by the China Scholarship Council through the Chinese Government Graduate Student Overseas Study Program.
The authors also thank Pradipta Mukherjee (University of Tsukuba) for providng careful review of the manuscript.

\section*{Funding}
Core Research for Evolutional Science and Technology (JPMJCR2105);
Japan Science and Technology Agency (JPMJMI18G8);
Japan Society for the Promotion of Science (18H01893, 21H01836, 22K04962);
China Scholarship Council (201908130130);
Austrian Science Fund (J4460, Schr\"odinger grand).

\section*{Disclosures}
L. Zhu, Makita, Lichtenegger, Lim, Y. Zhu, Yasuno: Yokogawa Electric Corp. (F), Sky technology (F), Nikon (F), Kao Corp. (F), Topcon (F).
Tamaoki, Kobayashi: None.

\section*{Data Availability}
Data underlying the results presented in this paper are not publicly available at this time but may be obtained from the authors upon reasonable request. 

\label{sec:refs}
\bibliography{reference.bib}

\end{document}